\documentclass[aps,prl,
reprint,
letterpaper,
groupedaddress,
twocolumn,
floatfix,
amsmath,amssymb,amsfonts,
]{revtex4-1}

\usepackage{graphicx}
\usepackage[breaklinks=true]{hyperref}
\usepackage{color}
\usepackage{array}
\usepackage[normalem]{ulem}

\newcolumntype{C}[1]{>{\centering\let\newline\\\arraybackslash\hspace{0pt}}m{#1}}

\begin{document}

\title{Towards an optimal flow: Density-of-states-informed replica-exchange simulations}

\author{Thomas Vogel}
\email{tvogel@lanl.gov}
\thanks{Current address: Department of Phys\-ics, Stetson University, DeLand, FL 32723}
\author{Danny Perez}
\email{danny\_perez@lanl.gov}

\affiliation{Theoretical Division (T-1), Los Alamos National Laboratory, Los Alamos, NM 87545}

\begin{abstract}
  Replica exchange (RE) is one of the most popular enhanced-sampling
  simulations technique in use today. Despite widespread successes, RE
  simulations can sometimes fail to converge in practical amounts of
  time, e.g., when sampling around phase transitions, or when a few
  hard-to-find configurations dominate the statistical averages. We
  introduce a generalized RE scheme, density-of-states-informed RE
  (g-RE), that addresses some of these challenges. The key feature of
  our approach is to inform the simulation with readily available, but
  commonly unused, information on the the density of states of the
  system as the RE simulation proceeds. This enables two improvements,
  namely, the introduction of resampling moves that actively move the
  system towards equilibrium, and the continual adaptation of the
  optimal temperature set. As a consequence of these two innovations,
  we show that the configuration flow in temperature space is
  optimized and that the overall convergence of RE simulations can be
  dramatically accelerated.
\end{abstract}

\maketitle

Sampling the phase space of Hamiltonians to estimate thermodynamic
properties is one of the fundamental problems in statistical physics.
However, direct approaches often dramatically fail due to the presence
of large free-energy barriers between different regions of phase
space. While many methods have been proposed to address this
challenge, few have had an impact as significant as the replica
exchange (RE)
method~\cite{geyer91proc,lyubartsev92jcp,hukushima96jpsj,earl05pccp}.
Indeed, since its introduction RE has established itself as a
workhorse in atomistic and coarse-grained simulations and is currently
used to investigate a large variety of systems in many areas ranging
from statistical
physics~\cite{hasenbusch08prb,yucesoy12prl,bittner06epl}, over biology
and chemistry~\cite{hansmann97cpl,gross13jcp,tsai05pnas}, to
solid-state physics and materials
science~\cite{falcioni99jcp,auer01nature,chuang05surfsci}.

RE enhances the exploration of phase space by using a set of $N$
individual simulations (often called ``walkers'') that evolve through
Monte Carlo (MC) or molecular dynamics (MD) updates under different
external parameters. For example, each walker might run at a different
temperature $T_i$ ($1\leq i \leq N$), which is the situation we
consider in the following (in this context, RE is often referred to as
Parallel Tempering). After predefined time intervals
$\tau_\mathrm{RE}$, the exchange of the current microstates between
pairs of walkers is attempted and carried out with an appropriate
probability;
$W_\mathrm{acc}(U_i,U_j)=\min[1,\mathrm{e}^{\Delta\beta\,\Delta U}]$
in the case of canonical walkers, where
$\beta_i=(k_\mathrm{B}T_i)^{-1}$ are the inverse temperatures of the
heat baths and $U_i$ the internal potential energies of the
microstates.  This exchange mechanism promotes configurational mixing
by exposing replicas to external conditions (e.g., high temperatures)
where free energy barriers can easily be overcome. It further provides
a means for thermodynamic information to be transferred to conditions
where the convergence of direct MC or MD simulations would require
prohibitively long simulation times.

The exchange probabilities in RE strictly comply with detail balance,
which ensures that the proper canonical distributions will be sampled
at all temperatures. This property is extremely useful, because
samples taken at each temperatures can be used without reweighting. It
can however be restrictive, especially in the early stages of a
simulation, when systems are far from equilibrium. This is related to
the fact that conventional RE does not provide a natural mechanism to
integrate and exploit information that becomes available during the
simulation.

In this manuscript, we show how one such inexpensive and readily
available source of information, namely, concurrent estimates of the
density of states, $g(U)$, can significantly improve conventional RE.
We leverage the (approximate) knowledge of $g(U)$ in two ways: first,
we introduce a resampling operation akin to a Gibbs sampling move,
which samples according to the $g(U)$-inferred canonical distribution
over an ensemble of configurations previously visited by any replica;
as we will show thereby explicitly steering the system toward
equilibrium.  Second, we use estimates of $g(U)$ to continuously
improve the temperature set $\{T_i\}$.  Because resampling breaks
correlations along individual trajectories on each replica, the
temperature set can be made optimal with respect to diffusion in
temperature space~\cite{nadler07pre}.

The enabling factor of our approach is the concurrent estimation of
the density of states $g(U)$. While it can be obtained by a number of
techniques~\cite{wang01prl,laio02pnas,kim06prl,junghans14jctc}, we
here rely on ideas from the Adaptive Biasing Force (ABF)
formalism~\cite{darve08jcp} for MD simulations.  The key is to frame
the problem as the estimation of the free energy
$F_\beta(U)=-k_\mathrm{B}T\ln [g(U)\,\exp(-\beta U)]$ through its
derivative $s$, which can be written in terms of microcanonical
averages~\cite{darve08jcp} as:
\begin{equation}
\label{eq_10}
s(U)=\frac{\mathrm{d}F}{\mathrm{d}U}=-\left\langle \frac{\mathrm{d}}{\mathrm{d}t}(w\cdot p)\right\rangle_U\,, 
\end{equation}
with $p$ being the vector of momenta and $w=\nabla U/(\nabla
U\cdot\nabla U)$. Here, the derivative with respect to time is
understood as a derivative along a micro-canonical trajectory. In
practice, $\mathrm{d}(w\cdot p)/{\mathrm{d}t}$ is measured
periodically (say every 100 timesteps) and stored with its
corresponding value of $U$.  We then use binned averages to
reconstruct $s(U)$. By integration, we recover $F_\beta(U)$, and hence
$g(U)$~\footnote{For brevity, we will use the same symbol for the
  exact $g(U)$ and its estimator in the following.  It should however
  be understood that, in the practical implementation of g-RE, the
  exact $g(U)$ is not available, so that $g(U)$ there refers to the
  estimator based on Eq.~(\ref{eq_10}) instead.}. In an alternative
representation, $s(U)$ can be used to define a microcanonical
temperature observable: $T_\mathrm{m}(U)=T_0/(1-s(U))$, where $T_0$ is
the heat-bath temperature. By integrating the thermodynamic relation
$1/T_\mathrm{m}(U)=\mathrm{d}S(U)/\mathrm{d}U$, with
$S(U)=k_\mathrm{B}\ln g(U)$, one reaches the same result.  Note that
if the momenta are unavailable, e.g., when using MC dynamics, a
configurational temperature $T_\mathrm{m}(U)$ can be estimated based
on structural data alone~\cite{rugh97prl,butler98jcp,SM}.  The cost of
estimating $s(U)$ is negligible in practice. An advantage of this
approach is that the validity of Eq.~(\ref{eq_10}) does not depend on
the sampling being carried out in any particular ensemble: the only
requirement is that the dynamics yield equal probabilities of
observing different configurations with the same $U$.  To minimize the
impact of initial --- potentially far-from-equilibrium --- states on
the estimator at later times, we introduce a memory time (much larger
than all other time scales) after which measurements are discarded.

While in the ABF method~\cite{darve08jcp}, $s(U)$ is used to create a
multicanonical ensemble, we instead leverage it in \hbox{g-RE} in the
following ways. While leaving the original RE mechanism untouched, we
first introduce an additional, global resampling move (executed after
time intervals $\tau_{\mathrm{resamp}}$) by which the microstate of a
replica is resampled from the ensemble of configurations visited by
{\em any} of its peers at {\em any} time in the past. In practice,
this is enabled by a global configuration database populated by all
walkers during the simulation. A configuration is selected from the
database with a probability proportional to its estimated canonical
weight $P_{\beta_i}(U)=g(U)\exp[-\beta_iU]$. This is akin to an
approximate Gibbs sampling~\cite{liu02book,chodera11jcp}. What is
crucial here is that the $P_{\beta_i}(U)$ are inferred from {\em
  global} thermodynamic information, so that it can differ from the
distribution locally sampled by the corresponding walker.  In other
words, this operation actively steers the distributions towards what
is globally deemed equilibrium. It does so by allowing for the {\em
  replication} of thermodynamically relevant states, in contrast to
conventional RE where configuration can only {\em diffuse} in
configuration space. As will be shown below, resampling proves
essential for convergence in the neighborhood of strong phase
transitions and for the timely escape out of metastable states. {One
  might, however, wonder whether g-RE produces correct statistics, as
  resampling does not {\em a priori} obey global balance when the
  configuration database is finite. In fact, the only requirement for
  correctness is that any two state with the same $U$ have the same
  probability of eventually being observed during an arbitrary long
  simulation.
  
  As discussed in the Supplementary Materials below, this remains the
  case when resampling is introduced, as long as the samplers on each
  replicas (e.g., Langevin-MD or Metropolis MC) are ergodic and
  canonical in and of themselves.  In that limit, $g(U)$, as obtained
  through Eq.~\ref{eq_10} will converge to the correct value.  From
  the knowledge of $g(U)$ and from a sample of observed
  configurations, any canonical quantity can then be obtained.  It is
  however important to note that resampling introduces correlations
  between the configurations stored in the database at a specific
  point in time, and hence, potentially also between replicas. Our
  approach exploits these correlations to share information between
  replicas.  However, resampling too frequently will lead to
  statistical ineficiencies. The optimal choice of
  $\tau_{\mathrm{resamp}}$ will be discussed in future works.
\begin{figure*}[t]
\includegraphics[width=.88\textwidth]{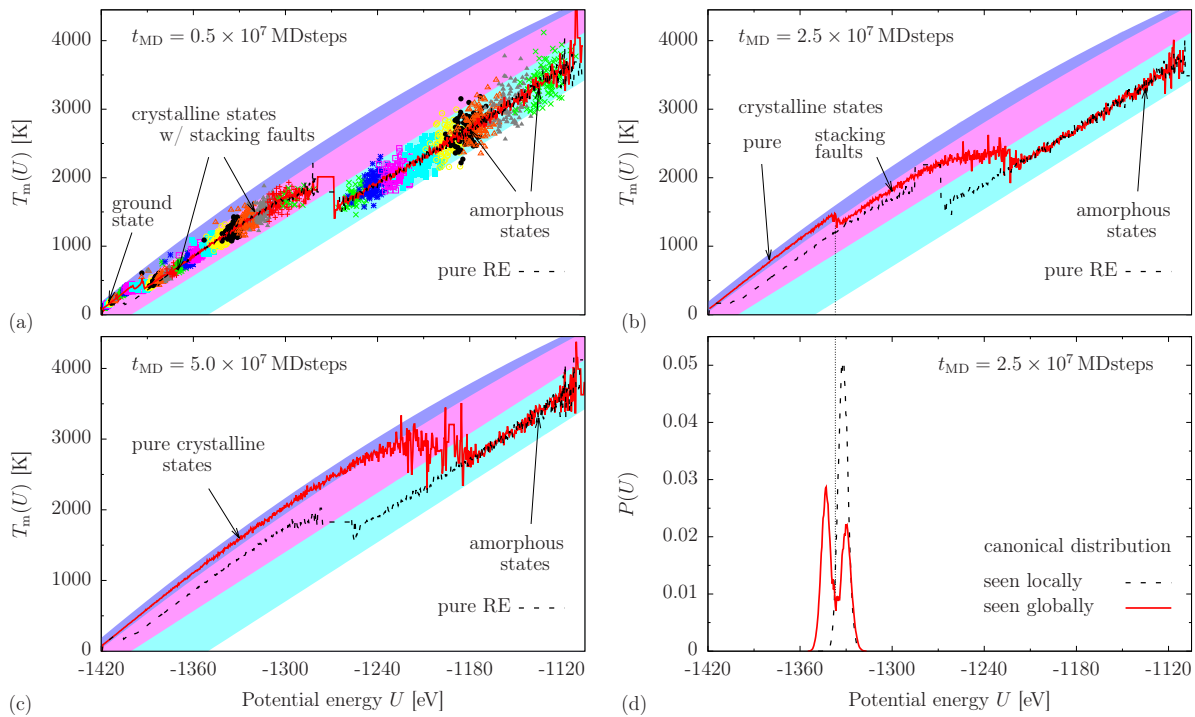}
\caption{Convergence of $T_\mathrm{m}(U)$ at different simulation
  times: (a) shortly after the start of the run and after the first
  pure crystalline states have been discovered; (b) these states get
  replicated and move the spurious transition between pure and faulted
  states upwards in temperature; (c) after convergence of the g-RE
  scheme.  The dots in (a) exemplarily show individual measurements,
  different colors correspond to different walkers (only a subset of
  data from every other walker is shown). The solid, red line shows
  $T_\mathrm{m}(U)$, the dashed black line the same obtained from an
  equivalent, conventional RE run for comparison. The shaded bands in
  panels (a) to (c) are guides to the eyes and denote the type of
  configurations predominantly found in each of these regions, as
  indicated by the different labels. Panel (d) shows the estimated canonical
  distribution $P_{\beta_i}(U)$, at the time corresponding to panel (b),
 as obtained from global data (solid, red line) compared to
  what would be inferred by an isolated walker trapped in
  faulted states (dotted, black line).
  \label{fig_2}}
\end{figure*}

The availability of $g(U)$ also enables a second innovation: the
continuous optimization of the temperature set $\{T_i\}$.  The overall
goal here is to minimize the round-trip time for replicas to wander
between low and high temperatures.  Much effort has been (and is
still) dedicated to addressing this issue (see
Refs.~\cite{rathore05jcp,katzgraber06jsm,trebst06jcp,nadler07pre,neuhaus07pre,patriksson08pccp,bittner08prl,ballard14jctc}
for examples). From this body of work it emerges that performance is
often characterized in terms of two key concepts: the average exchange
acceptance probabilities $W_\mathrm{acc}(i,i+1)$ between pairs of
neighboring temperatures, and the flow ratio
$f(i)=n_\mathrm{up}(i)/[n_\mathrm{up}(i)+n_\mathrm{down}(i)]$, i.e.,
the fraction of replicas that diffuse up in temperature for a given
walker~$i$ (a replica is said to flow upwards if it visited the
minimum temperature more recently than the maximum temperature,
cf.~\cite{katzgraber06jsm,trebst06jcp}). In the probability-centric
view~\cite{nadler07pre,neuhaus07pre,bittner08prl}, the objective is to
find the $\{T_i\}$ such that
$W_\mathrm{acc}(i,i+1)=\mathrm{const},\,\forall i$, the insight being
that locally low acceptance probabilities would limit the free
diffusion of replicas.  In the flow-centric
view~\cite{katzgraber06jsm,trebst06jcp}, the optimal set $\{T_i\}$ is
such that $f(i)=1-[(i-1)/(N-1)]$ (for $T_i<T_{i+1},\,\forall i$), as
this indicates an unimpeded flow of replicas.  In contrast, the
presence of a bottleneck would signal itself by comparatively flat
regions separated by a sharp drop in $f(i)$. The two approaches have
been contrasted by Nadler and Hansmann~\cite{nadler07pre} who showed
that, in the special case where the dynamics on each replica become
completely uncorrelated between exchange attempts, the {\em optimal}
choice is to make the exchange probabilities constant, as this
minimizes the round-trip time from the lowest to the highest
temperatures. Furthermore, the flow ratio will also be linear in this
case. In the current context, this limit can be approached by setting
$\tau_{\mathrm{resamp}}=\tau_\mathrm{RE}$, as resampling decreases
correlations (as long as $\tau_{RE}$ is not so short as to saturate
the database with near-identical configurations).  In g-RE, the
optimal temperature set can easily be determined.  Holding the minimal
and the maximal temperatures fixed, we apply a bisection scheme until
$\{T_i\}$ converges to a situation where
\begin{align}
\nonumber
W^\mathrm{m}_\mathrm{acc}(P_{i+1}, P_i)=\int_{-\infty}^\infty&
P_{\beta_i}(U)\int_{-\infty}^\infty W_\mathrm{acc}(U,U^\prime)\\ \label{eq_12}
&P_{\beta_{i+1}}(U^\prime) \, \mathrm{d}U \mathrm{d}U^\prime=\mathrm{const},\; \forall i\,.
\end{align}
(see Supplemental Materials below for more details). Here also,
the $P_{\beta_i}(U)$ are based on the current estimator of $g(U)$, and
Eq.~(\ref{eq_12}) can be evaluated numerically.  This scheme is very
fast in practice and does not require any preliminary calculation. We
continuously readjust the temperatures at intervals
$\tau_{\mathrm{adapt}}$, assisting convergence in situations where the
walkers begin far from equilibrium. Note that adaption of the
$\{T_i\}$ does not interfere with the averages necessary for the
measurement of $s(U)$, as samples taken at different temperatures can
be seamlessly integrated through Eq.~(\ref{eq_10}).

We now demonstrate the performance of our method on a system of 500
silver atoms. The $N=71$ walkers are canonical molecular dynamics runs
with a timestep of $\tau_\mathrm{ts}=2$\,fs and the heat-bath
temperatures are set by Langevin thermostats.  Atoms interact via an
embedded-atom potential~\cite{williams06msmse}. The simulation cell is
cubic and periodic boundary conditions are used in all directions.
The minimum temperature is set to $T_\mathrm{min}=100$\,K and the
maximal to $T_\mathrm{max}=3500$\,K.  The particle density is fixed at
$\rho=0.0585$\,\AA$^{-3}$, which corresponds to the density that
minimizes the energy of the fcc crystal; the fcc configuration is
therefore the putative global energy minimum for this system and
should dominate up to the vicinity of the melting point, given that
the thermal concentration of defects is expected to be vanishingly
small. However, all walkers are initialized from a quenched liquid
(amorphous) configuration. This choice makes the system a very good
prototype of a case where the thermodynamically relevant
configurations are unknown {\em a priori} and difficult to access.
Indeed, from MD studies of metals (e.g., Ag
\cite{xiao06jcp,Tian08ncs,jian10scts}, Cu \cite{liu01jcp}, Ni
\cite{li2011int}, etc.) it is known that recovering the perfect
crystalline state from the melt requires very slow cooling, otherwise
the system remains trapped in amorphous states (at fast cooling) or in
a mixture of fcc and hcp regions separated by stacking defects (at
moderate cooling rates).  In addition, the presence of a first-order
transition (melting) within the range of temperatures makes this an
extremely challenging system to study. Finally, since the thermal
distributions of crystal defects is expected to be vanishingly small
in a system of that size away from the immediate vicinity of the
melting point, the validity of the results is easy to assess.

As illustrated in Fig.~\ref{fig_2}, the coupling of the different
replicas through the estimator of $g(U)$ and the configurational
database directs the evolution of the system towards thermal
equilibrium by enabling the {\em replication} of thermodynamically
relevant, crystalline states, in contrast to conventional RE where
states are only {\em exchanged}.  In that figure, we report the
instantaneous estimators of the microcanonical temperature
$T_\mathrm{m}(U)$ at different times in the simulation. In our
example, the perfect crystalline states (upper, purple band) are
essentially the only relevant ones below the melting point; they are,
however, by construction not present in the early stages of the
simulations.  While, at lower temperatures, the system quickly leaves
the amorphous region of phase space (lower, pale blue band) and
crystallizes, most of these crystalline configurations initially
contain stacking faults (middle, pink band) that take a very long time
to anneal.  Convergence of standard RE requires the perfect
crystalline state to be {\em independently} found at least (and
ideally much more than) $n_{<melt}$ times (the number of walkers for
which $T_i<T_\mathrm{melt}$).  The rate at which this occurs hence
controls the convergence speed; from standard RE we infer this rate
to be of the order $10^8\,\mathrm{s}^{-1}$ and the convergence would
require approximately $2.5 \times 10^8$ MD steps. In contrast,
resampling enables the replication of the pure crystalline state as
soon at it is found {\em once} by {\em any} replica.  As illustrated
in Fig.~\ref{fig_2}d, this occurs because the {\em global} estimate of
$g(U)$ eventually contains contributions from the crystalline region
that are locally invisible to a walker trapped in faulted states.
Computational gains follow because the probability of resampling a
crystalline state significantly exceeds that of a replica
independently finding it.

The same is true around the melting transition, which is initially
biased to lower temperatures due to the initialization from a quenched
liquid (note that the melting point at constant volume is much higher
than the triple point). Before equilibration, a spurious transition
between perfect crystals and faulted states (the sudden drop in
$T_\mathrm{m}(U)$ at low energies) is observed and the melting
temperature is underestimated.  This is illustrated in
Fig.~\ref{fig_2}\,a--b (red, solid lines). A particular advantage of
continuous temperature adaptation is that the temperature set remains
nearly optimal at all times, even as the position of the (spurious or
real) phase transitions evolve during convergence. With g-RE we
observe convergence at $t_\mathrm{MD}\approx 4\times10^7$\,MD steps,
while conventional RE, even with temperature adaptation (black, dashed
lines), still mainly samples faulted states. The temperature
adaptation scheme is very robust and converges quickly, even for a
poor choice of the initial set $\{T_i\}$ (see below).

\begin{figure}
\includegraphics[width=\columnwidth]{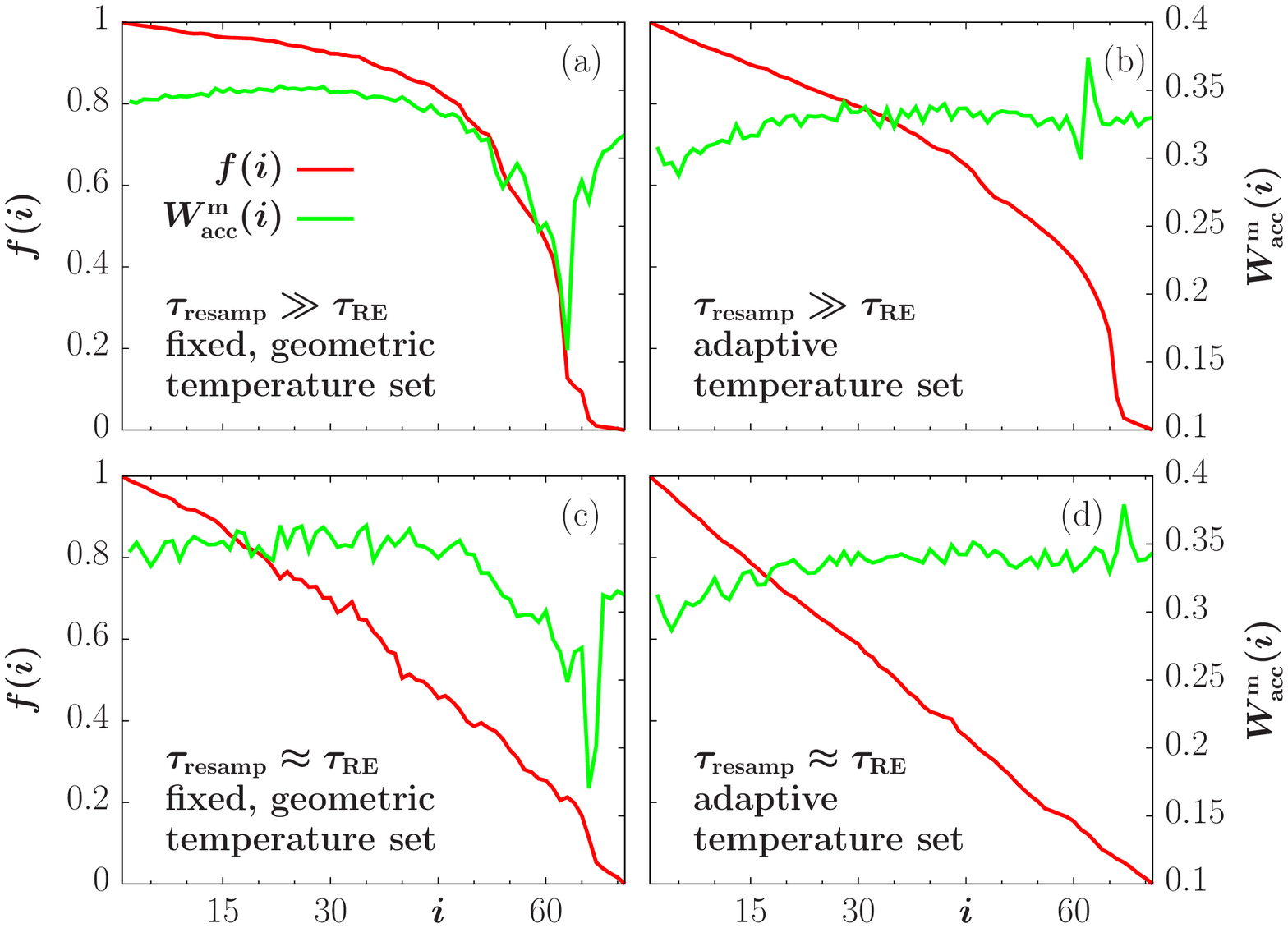}
\caption{Measured RE acceptance rates ($W_\mathrm{acc}^\mathrm{m}(i,i+1)$;
  green curves) and fraction of replicas diffusing from the lowest to
  the highest temperature ($f(i)$; red curves) for the 500 Ag atoms system. (a,c) fixed, geometric
  temperature set (b,d) adaptive temperature set; (a,b) non-ergodic
  sampler: $\tau_{\mathrm{resamp}}\gg\tau_{\mathrm{RE}}$, (c,d)
  ergodic sampler: $\tau_{\mathrm{resamp}}\approx\tau_{\mathrm{RE}}$.
 \label{fig_1}
}
\end{figure}
Finally, Fig.~\ref{fig_1} shows how the introduction of global resampling
at time scales comparable to the RE exchange time
($\tau_\mathrm{resamp}\approx\tau_\mathrm{RE}\sim
10^2\tau_\mathrm{ts}$) and an adaptive temperature set results in an
optimal flow of the replica through temperature space, i.e., {\em
  both} constant $W_\mathrm{acc}^\mathrm{m}(i,i+1)$ {\em and} linear
$f(i)$ are observed (Fig.~\ref{fig_1}\,d). 
The measured round-trip times
($\tau_\mathrm{rt}^\mathrm{m}\approx14000\,\tau_\mathrm{RE}$) are in
perfect agreement with measured round-trip times for a purely random
exchange process with corresponding exchange probabilities, which here
correspond to the optimal limit~\cite{nadler07pre}. As show in
Fig.~\ref{fig_1}\,a-c, these two conditions can only be obeyed when using
both temperature adaptation and resampling.
{In this sense, our work integrates earlier efforts ~\cite{bittner08prl,trebst06jcp,katzgraber06jsm}, where
  either constant exchange times, exchange rates, or the optimal flow
  had to be sacrificed.}

In conclusion, we introduce a general scheme, g-RE, that can
dramatically improve the efficiency of RE simulations.  The method is
based on the idea of informing RE simulations with estimators of the
density of states $g(U)$ gathered on the fly. This allows for two key
improvements: the introduction of a global resampling move that guides
the system towards equilibrium and causes a dramatic reductions of
correlation times of the sampling, and the on-the-fly determination of
an optimal temperature set that simultaneously achieves constant
exchange probabilities and linear flow ratio.  We expect our method to
be particularly useful for any system with dominant but hard to access
states or around strong first-order transitions. So far, there have
been two approaches to such problems: either to change the ensemble
(see~\cite{neuhaus07pre,kim10jcp} for examples) or to introduce global
MC trial moves. In the former case, one often chooses to work in the
multicanonical ensemble where $P(U)=\mathrm{const}$; creating such an
ensemble is in fact a common way to leverage the knowledge of $g(U)$.
This approach, however, is not optimal for the system reported here,
because even though that ensemble is free from gradients of $F(U)$,
the energy landscape locally remains extremely rough, thereby severely
limiting the diffusivity in $U$.  Using diffusion in $T$-space to
promote mixing proved a more efficient alternative. Nonetheless, it
still required the introduction of a global move to insure
convergence. We here proposed a {\em generic} solution for
constructing such a global move that does not require any {\em a
  priori } information about the system. We hence expect this approach
to be useful for a wide range of systems and to also be applicable to
RE schemes using other
ensembles~\cite{gross13jcp,sugita00cpl,kim12jpcb,vogel13prl,vogel14pre,auer01nature}.

\begin{acknowledgments}
  This work was supported by the U.S. Department of Energy through the
  Los Alamos National Laboratory (LANL)/LDRD Program and used
  computing resources provided by the Los Alamos National Laboratory
  Institutional Computing Program.  Los Alamos National Laboratory is
  operated by Los Alamos National Security, LLC, for the National
  Nuclear Security administration of the US DOE under contract
  DE-AC52-06NA25396. LA-UR-15-23605 assigned.
\end{acknowledgments}

\bibliography{Gibbs_REMD.bib}

\begin{thebibliography}{47}%
\makeatletter
\providecommand \@ifxundefined [1]{%
 \@ifx{#1\undefined}
}%
\providecommand \@ifnum [1]{%
 \ifnum #1\expandafter \@firstoftwo
 \else \expandafter \@secondoftwo
 \fi
}%
\providecommand \@ifx [1]{%
 \ifx #1\expandafter \@firstoftwo
 \else \expandafter \@secondoftwo
 \fi
}%
\providecommand \natexlab [1]{#1}%
\providecommand \enquote  [1]{``#1''}%
\providecommand \bibnamefont  [1]{#1}%
\providecommand \bibfnamefont [1]{#1}%
\providecommand \citenamefont [1]{#1}%
\providecommand \href@noop [0]{\@secondoftwo}%
\providecommand \href [0]{\begingroup \@sanitize@url \@href}%
\providecommand \@href[1]{\@@startlink{#1}\@@href}%
\providecommand \@@href[1]{\endgroup#1\@@endlink}%
\providecommand \@sanitize@url [0]{\catcode `\\12\catcode `\$12\catcode
  `\&12\catcode `\#12\catcode `\^12\catcode `\_12\catcode `\%12\relax}%
\providecommand \@@startlink[1]{}%
\providecommand \@@endlink[0]{}%
\providecommand \url  [0]{\begingroup\@sanitize@url \@url }%
\providecommand \@url [1]{\endgroup\@href {#1}{\urlprefix }}%
\providecommand \urlprefix  [0]{URL }%
\providecommand \Eprint [0]{\href }%
\providecommand \doibase [0]{http://dx.doi.org/}%
\providecommand \selectlanguage [0]{\@gobble}%
\providecommand \bibinfo  [0]{\@secondoftwo}%
\providecommand \bibfield  [0]{\@secondoftwo}%
\providecommand \translation [1]{[#1]}%
\providecommand \BibitemOpen [0]{}%
\providecommand \bibitemStop [0]{}%
\providecommand \bibitemNoStop [0]{.\EOS\space}%
\providecommand \EOS [0]{\spacefactor3000\relax}%
\providecommand \BibitemShut  [1]{\csname bibitem#1\endcsname}%
\let\auto@bib@innerbib\@empty
\bibitem [{\citenamefont {Geyer}(1991)}]{geyer91proc}%
  \BibitemOpen
  \bibfield  {author} {\bibinfo {author} {\bibfnamefont {C.~J.}\ \bibnamefont
  {Geyer}},\ }in\ \href@noop {} {\emph {\bibinfo {booktitle} {Computing Science
  and Statistics: Proceedings of the 23rd Symposium on the Interface}}},\
  \bibinfo {editor} {edited by\ \bibinfo {editor} {\bibfnamefont {E.~M.}\
  \bibnamefont {Keramidas}}}\ (\bibinfo  {publisher} {Interface Foundation},\
  \bibinfo {address} {Fairfax Station, VA},\ \bibinfo {year} {1991})\ p.\
  \bibinfo {pages} {156}\BibitemShut {NoStop}%
\bibitem [{\citenamefont {Lyubartsev}\ \emph {et~al.}(1992)\citenamefont
  {Lyubartsev}, \citenamefont {Martsinovski}, \citenamefont {Shevkunov},\ and\
  \citenamefont {Vorontsov-Velyaminov}}]{lyubartsev92jcp}%
  \BibitemOpen
  \bibfield  {author} {\bibinfo {author} {\bibfnamefont {A.~P.}\ \bibnamefont
  {Lyubartsev}}, \bibinfo {author} {\bibfnamefont {A.~A.}\ \bibnamefont
  {Martsinovski}}, \bibinfo {author} {\bibfnamefont {S.~V.}\ \bibnamefont
  {Shevkunov}}, \ and\ \bibinfo {author} {\bibfnamefont {P.~N.}\ \bibnamefont
  {Vorontsov-Velyaminov}},\ }\href {\doibase 10.1063/1.462133} {\bibfield
  {journal} {\bibinfo  {journal} {J. Chem. Phys.}\ }\textbf {\bibinfo {volume}
  {96}},\ \bibinfo {pages} {1776} (\bibinfo {year} {1992})}\BibitemShut
  {NoStop}%
\bibitem [{\citenamefont {Hukushima}\ and\ \citenamefont
  {Nemoto}(1996)}]{hukushima96jpsj}%
  \BibitemOpen
  \bibfield  {author} {\bibinfo {author} {\bibfnamefont {K.}~\bibnamefont
  {Hukushima}}\ and\ \bibinfo {author} {\bibfnamefont {K.}~\bibnamefont
  {Nemoto}},\ }\href {\doibase 10.1143/JPSJ.65.1604} {\bibfield  {journal}
  {\bibinfo  {journal} {J. Phys. Soc. Japan}\ }\textbf {\bibinfo {volume}
  {65}},\ \bibinfo {pages} {1604} (\bibinfo {year} {1996})}\BibitemShut
  {NoStop}%
\bibitem [{\citenamefont {Earl}\ and\ \citenamefont {Deem}(2005)}]{earl05pccp}%
  \BibitemOpen
  \bibfield  {author} {\bibinfo {author} {\bibfnamefont {D.~J.}\ \bibnamefont
  {Earl}}\ and\ \bibinfo {author} {\bibfnamefont {M.~W.}\ \bibnamefont
  {Deem}},\ }\href {\doibase 10.1039/B509983H} {\bibfield  {journal} {\bibinfo
  {journal} {Phys. Chem. Chem. Phys.}\ }\textbf {\bibinfo {volume} {7}},\
  \bibinfo {pages} {3910} (\bibinfo {year} {2005})}\BibitemShut {NoStop}%
\bibitem [{\citenamefont {Hasenbusch}\ \emph {et~al.}(2008)\citenamefont
  {Hasenbusch}, \citenamefont {Pelissetto},\ and\ \citenamefont
  {Vicari}}]{hasenbusch08prb}%
  \BibitemOpen
  \bibfield  {author} {\bibinfo {author} {\bibfnamefont {M.}~\bibnamefont
  {Hasenbusch}}, \bibinfo {author} {\bibfnamefont {A.}~\bibnamefont
  {Pelissetto}}, \ and\ \bibinfo {author} {\bibfnamefont {E.}~\bibnamefont
  {Vicari}},\ }\href {\doibase 10.1103/PhysRevB.78.214205} {\bibfield
  {journal} {\bibinfo  {journal} {Phys. Rev. B}\ }\textbf {\bibinfo {volume}
  {78}},\ \bibinfo {pages} {214205} (\bibinfo {year} {2008})}\BibitemShut
  {NoStop}%
\bibitem [{\citenamefont {Yucesoy}\ \emph {et~al.}(2012)\citenamefont
  {Yucesoy}, \citenamefont {Katzgraber},\ and\ \citenamefont
  {Machta}}]{yucesoy12prl}%
  \BibitemOpen
  \bibfield  {author} {\bibinfo {author} {\bibfnamefont {B.}~\bibnamefont
  {Yucesoy}}, \bibinfo {author} {\bibfnamefont {H.~G.}\ \bibnamefont
  {Katzgraber}}, \ and\ \bibinfo {author} {\bibfnamefont {J.}~\bibnamefont
  {Machta}},\ }\href {\doibase 10.1103/PhysRevLett.109.177204} {\bibfield
  {journal} {\bibinfo  {journal} {Phys. Rev. Lett.}\ }\textbf {\bibinfo
  {volume} {109}},\ \bibinfo {pages} {177204} (\bibinfo {year}
  {2012})}\BibitemShut {NoStop}%
\bibitem [{\citenamefont {Bittner}\ and\ \citenamefont
  {Janke}(2006)}]{bittner06epl}%
  \BibitemOpen
  \bibfield  {author} {\bibinfo {author} {\bibfnamefont {E.}~\bibnamefont
  {Bittner}}\ and\ \bibinfo {author} {\bibfnamefont {W.}~\bibnamefont
  {Janke}},\ }\href {\doibase 10.1209/epl/i2006-10007-y} {\bibfield  {journal}
  {\bibinfo  {journal} {EPL (Europhys. Lett.)}\ }\textbf {\bibinfo {volume}
  {74}},\ \bibinfo {pages} {195} (\bibinfo {year} {2006})}\BibitemShut
  {NoStop}%
\bibitem [{\citenamefont {Hansmann}(1997)}]{hansmann97cpl}%
  \BibitemOpen
  \bibfield  {author} {\bibinfo {author} {\bibfnamefont {U.~H.~E.}\
  \bibnamefont {Hansmann}},\ }\href {\doibase 10.1016/S0009-2614(97)01198-6}
  {\bibfield  {journal} {\bibinfo  {journal} {Chem. Phys. Lett.}\ }\textbf
  {\bibinfo {volume} {281}},\ \bibinfo {pages} {140} (\bibinfo {year}
  {1997})}\BibitemShut {NoStop}%
\bibitem [{\citenamefont {Gross}\ \emph {et~al.}(2013)\citenamefont {Gross},
  \citenamefont {Neuhaus}, \citenamefont {Vogel},\ and\ \citenamefont
  {Bachmann}}]{gross13jcp}%
  \BibitemOpen
  \bibfield  {author} {\bibinfo {author} {\bibfnamefont {J.}~\bibnamefont
  {Gross}}, \bibinfo {author} {\bibfnamefont {T.}~\bibnamefont {Neuhaus}},
  \bibinfo {author} {\bibfnamefont {T.}~\bibnamefont {Vogel}}, \ and\ \bibinfo
  {author} {\bibfnamefont {M.}~\bibnamefont {Bachmann}},\ }\href {\doibase
  10.1063/1.4790615} {\bibfield  {journal} {\bibinfo  {journal} {J. Chem.
  Phys.}\ }\textbf {\bibinfo {volume} {138}},\ \bibinfo {pages} {074905}
  (\bibinfo {year} {2013})}\BibitemShut {NoStop}%
\bibitem [{\citenamefont {Tsai}\ \emph {et~al.}(2005)\citenamefont {Tsai},
  \citenamefont {Reches}, \citenamefont {Tsai}, \citenamefont {Gunasekaran},
  \citenamefont {Gazit},\ and\ \citenamefont {Nussinov}}]{tsai05pnas}%
  \BibitemOpen
  \bibfield  {author} {\bibinfo {author} {\bibfnamefont {H.-H.~G.}\
  \bibnamefont {Tsai}}, \bibinfo {author} {\bibfnamefont {M.}~\bibnamefont
  {Reches}}, \bibinfo {author} {\bibfnamefont {C.-J.}\ \bibnamefont {Tsai}},
  \bibinfo {author} {\bibfnamefont {K.}~\bibnamefont {Gunasekaran}}, \bibinfo
  {author} {\bibfnamefont {E.}~\bibnamefont {Gazit}}, \ and\ \bibinfo {author}
  {\bibfnamefont {R.}~\bibnamefont {Nussinov}},\ }\href {\doibase
  10.1073/pnas.0408653102} {\bibfield  {journal} {\bibinfo  {journal} {Proc.
  Natl. Acad. Sci. U.S.A.}\ }\textbf {\bibinfo {volume} {102}},\ \bibinfo
  {pages} {8174} (\bibinfo {year} {2005})}\BibitemShut {NoStop}%
\bibitem [{\citenamefont {Falcioni}\ and\ \citenamefont
  {Deem}(1999)}]{falcioni99jcp}%
  \BibitemOpen
  \bibfield  {author} {\bibinfo {author} {\bibfnamefont {M.}~\bibnamefont
  {Falcioni}}\ and\ \bibinfo {author} {\bibfnamefont {M.~W.}\ \bibnamefont
  {Deem}},\ }\href {\doibase 10.1063/1.477812} {\bibfield  {journal} {\bibinfo
  {journal} {J. Chem. Phys.}\ }\textbf {\bibinfo {volume} {110}},\ \bibinfo
  {pages} {1754} (\bibinfo {year} {1999})}\BibitemShut {NoStop}%
\bibitem [{\citenamefont {Auer}\ and\ \citenamefont
  {Frenkel}(2001)}]{auer01nature}%
  \BibitemOpen
  \bibfield  {author} {\bibinfo {author} {\bibfnamefont {S.}~\bibnamefont
  {Auer}}\ and\ \bibinfo {author} {\bibfnamefont {D.}~\bibnamefont {Frenkel}},\
  }\href {\doibase 10.1038/35059035} {\bibfield  {journal} {\bibinfo  {journal}
  {Nature}\ }\textbf {\bibinfo {volume} {409}},\ \bibinfo {pages} {1020}
  (\bibinfo {year} {2001})}\BibitemShut {NoStop}%
\bibitem [{\citenamefont {Chuang}\ \emph {et~al.}(2005)\citenamefont {Chuang},
  \citenamefont {Ciobanu}, \citenamefont {Predescu}, \citenamefont {Wang},\
  and\ \citenamefont {Ho}}]{chuang05surfsci}%
  \BibitemOpen
  \bibfield  {author} {\bibinfo {author} {\bibfnamefont {F.}~\bibnamefont
  {Chuang}}, \bibinfo {author} {\bibfnamefont {C.}~\bibnamefont {Ciobanu}},
  \bibinfo {author} {\bibfnamefont {C.}~\bibnamefont {Predescu}}, \bibinfo
  {author} {\bibfnamefont {C.}~\bibnamefont {Wang}}, \ and\ \bibinfo {author}
  {\bibfnamefont {K.}~\bibnamefont {Ho}},\ }\href {\doibase
  10.1016/j.susc.2005.01.036} {\bibfield  {journal} {\bibinfo  {journal} {Surf.
  Sci.}\ }\textbf {\bibinfo {volume} {578}},\ \bibinfo {pages} {183} (\bibinfo
  {year} {2005})}\BibitemShut {NoStop}%
\bibitem [{\citenamefont {Nadler}\ and\ \citenamefont
  {Hansmann}(2007)}]{nadler07pre}%
  \BibitemOpen
  \bibfield  {author} {\bibinfo {author} {\bibfnamefont {W.}~\bibnamefont
  {Nadler}}\ and\ \bibinfo {author} {\bibfnamefont {U.~H.~E.}\ \bibnamefont
  {Hansmann}},\ }\href {\doibase 10.1103/PhysRevE.75.026109} {\bibfield
  {journal} {\bibinfo  {journal} {Phys. Rev. E}\ }\textbf {\bibinfo {volume}
  {75}},\ \bibinfo {pages} {026109} (\bibinfo {year} {2007})}\BibitemShut
  {NoStop}%
\bibitem [{\citenamefont {Wang}\ and\ \citenamefont
  {Landau}(2001)}]{wang01prl}%
  \BibitemOpen
  \bibfield  {author} {\bibinfo {author} {\bibfnamefont {F.}~\bibnamefont
  {Wang}}\ and\ \bibinfo {author} {\bibfnamefont {D.~P.}\ \bibnamefont
  {Landau}},\ }\href {\doibase 10.1103/PhysRevLett.86.2050} {\bibfield
  {journal} {\bibinfo  {journal} {Phys. Rev. Lett.}\ }\textbf {\bibinfo
  {volume} {86}},\ \bibinfo {pages} {2050} (\bibinfo {year}
  {2001})}\BibitemShut {NoStop}%
\bibitem [{\citenamefont {Laio}\ and\ \citenamefont
  {Parrinello}(2002)}]{laio02pnas}%
  \BibitemOpen
  \bibfield  {author} {\bibinfo {author} {\bibfnamefont {A.}~\bibnamefont
  {Laio}}\ and\ \bibinfo {author} {\bibfnamefont {M.}~\bibnamefont
  {Parrinello}},\ }\href {\doibase 10.1073/pnas.202427399} {\bibfield
  {journal} {\bibinfo  {journal} {Proc. Natl. Acad. Sci.}\ }\textbf {\bibinfo
  {volume} {99}},\ \bibinfo {pages} {12562} (\bibinfo {year}
  {2002})}\BibitemShut {NoStop}%
\bibitem [{\citenamefont {Kim}\ \emph {et~al.}(2006)\citenamefont {Kim},
  \citenamefont {Straub},\ and\ \citenamefont {Keyes}}]{kim06prl}%
  \BibitemOpen
  \bibfield  {author} {\bibinfo {author} {\bibfnamefont {J.}~\bibnamefont
  {Kim}}, \bibinfo {author} {\bibfnamefont {J.~E.}\ \bibnamefont {Straub}}, \
  and\ \bibinfo {author} {\bibfnamefont {T.}~\bibnamefont {Keyes}},\ }\href
  {\doibase 10.1103/PhysRevLett.97.050601} {\bibfield  {journal} {\bibinfo
  {journal} {Phys. Rev. Lett.}\ }\textbf {\bibinfo {volume} {97}},\ \bibinfo
  {pages} {050601} (\bibinfo {year} {2006})}\BibitemShut {NoStop}%
\bibitem [{\citenamefont {Junghans}\ \emph {et~al.}(2014)\citenamefont
  {Junghans}, \citenamefont {Perez},\ and\ \citenamefont
  {Vogel}}]{junghans14jctc}%
  \BibitemOpen
  \bibfield  {author} {\bibinfo {author} {\bibfnamefont {C.}~\bibnamefont
  {Junghans}}, \bibinfo {author} {\bibfnamefont {D.}~\bibnamefont {Perez}}, \
  and\ \bibinfo {author} {\bibfnamefont {T.}~\bibnamefont {Vogel}},\ }\href
  {\doibase 10.1021/ct500077d} {\bibfield  {journal} {\bibinfo  {journal} {J.
  Chem. Theory Comput. (JCTC)}\ }\textbf {\bibinfo {volume} {10}},\ \bibinfo
  {pages} {1843} (\bibinfo {year} {2014})}\BibitemShut {NoStop}%
\bibitem [{\citenamefont {Darve}\ \emph {et~al.}(2008)\citenamefont {Darve},
  \citenamefont {Rodr{\'i}guez-G{\'o}mez},\ and\ \citenamefont
  {Pohorille}}]{darve08jcp}%
  \BibitemOpen
  \bibfield  {author} {\bibinfo {author} {\bibfnamefont {E.}~\bibnamefont
  {Darve}}, \bibinfo {author} {\bibfnamefont {D.}~\bibnamefont
  {Rodr{\'i}guez-G{\'o}mez}}, \ and\ \bibinfo {author} {\bibfnamefont
  {A.}~\bibnamefont {Pohorille}},\ }\href {\doibase 10.1063/1.2829861}
  {\bibfield  {journal} {\bibinfo  {journal} {J. Chem. Phys.}\ }\textbf
  {\bibinfo {volume} {128}},\ \bibinfo {pages} {144120} (\bibinfo {year}
  {2008})}\BibitemShut {NoStop}%
\bibitem [{Note1()}]{Note1}%
  \BibitemOpen
  \bibinfo {note} {For brevity, we will use the same symbol for the exact
  $g(U)$ and its estimator in the following. It should however be understood
  that, in the practical implementation of g-RE, the exact $g(U)$ is not
  available, so that $g(U)$ there refers to the estimator based on Eq.~(\ref
  {eq_10}) instead.}\BibitemShut {Stop}%
\bibitem [{\citenamefont {Rugh}(1997)}]{rugh97prl}%
  \BibitemOpen
  \bibfield  {author} {\bibinfo {author} {\bibfnamefont {H.~H.}\ \bibnamefont
  {Rugh}},\ }\href {\doibase 10.1103/PhysRevLett.78.772} {\bibfield  {journal}
  {\bibinfo  {journal} {Phys. Rev. Lett.}\ }\textbf {\bibinfo {volume} {78}},\
  \bibinfo {pages} {772} (\bibinfo {year} {1997})}\BibitemShut {NoStop}%
\bibitem [{\citenamefont {Butler}\ \emph {et~al.}(1998)\citenamefont {Butler},
  \citenamefont {Ayton}, \citenamefont {Jepps},\ and\ \citenamefont
  {Evans}}]{butler98jcp}%
  \BibitemOpen
  \bibfield  {author} {\bibinfo {author} {\bibfnamefont {B.~D.}\ \bibnamefont
  {Butler}}, \bibinfo {author} {\bibfnamefont {G.}~\bibnamefont {Ayton}},
  \bibinfo {author} {\bibfnamefont {O.~G.}\ \bibnamefont {Jepps}}, \ and\
  \bibinfo {author} {\bibfnamefont {D.~J.}\ \bibnamefont {Evans}},\ }\href
  {\doibase 10.1063/1.477301} {\bibfield  {journal} {\bibinfo  {journal} {J.
  Chem. Phys.}\ }\textbf {\bibinfo {volume} {109}},\ \bibinfo {pages} {6519}
  (\bibinfo {year} {1998})}\BibitemShut {NoStop}%
\bibitem [{SM()}]{SM}%
  \BibitemOpen
  \href@noop {} {}\bibinfo {note} {See Supplemental Material at [URL will be
  inserted by publisher], which includes
  Refs.~\cite{allen13molphys,junghans06prl,schnabel11pre,burden85book}}\BibitemShut
  {NoStop}%
\bibitem [{\citenamefont {Liu}(2002)}]{liu02book}%
  \BibitemOpen
  \bibfield  {author} {\bibinfo {author} {\bibfnamefont {J.~S.}\ \bibnamefont
  {Liu}},\ }\href@noop {} {\emph {\bibinfo {title} {Monte Carlo strategies in
  scientific computing}}},\ \bibinfo {edition} {2nd}\ ed.\ (\bibinfo
  {publisher} {Springer-Verlag},\ \bibinfo {address} {New York, NY},\ \bibinfo
  {year} {2002})\BibitemShut {NoStop}%
\bibitem [{\citenamefont {Chodera}\ and\ \citenamefont
  {Shirts}(2011)}]{chodera11jcp}%
  \BibitemOpen
  \bibfield  {author} {\bibinfo {author} {\bibfnamefont {J.~D.}\ \bibnamefont
  {Chodera}}\ and\ \bibinfo {author} {\bibfnamefont {M.~R.}\ \bibnamefont
  {Shirts}},\ }\href {\doibase 10.1063/1.3660669} {\bibfield  {journal}
  {\bibinfo  {journal} {J. Chem. Phys.}\ }\textbf {\bibinfo {volume} {135}},\
  \bibinfo {pages} {194110} (\bibinfo {year} {2011})}\BibitemShut {NoStop}%
\bibitem [{\citenamefont {Rathore}\ \emph {et~al.}(2005)\citenamefont
  {Rathore}, \citenamefont {Chopra},\ and\ \citenamefont
  {de~Pablo}}]{rathore05jcp}%
  \BibitemOpen
  \bibfield  {author} {\bibinfo {author} {\bibfnamefont {N.}~\bibnamefont
  {Rathore}}, \bibinfo {author} {\bibfnamefont {M.}~\bibnamefont {Chopra}}, \
  and\ \bibinfo {author} {\bibfnamefont {J.~J.}\ \bibnamefont {de~Pablo}},\
  }\href {\doibase 10.1063/1.1831273} {\bibfield  {journal} {\bibinfo
  {journal} {J. Chem. Phys.}\ }\textbf {\bibinfo {volume} {122}},\ \bibinfo
  {pages} {024111} (\bibinfo {year} {2005})}\BibitemShut {NoStop}%
\bibitem [{\citenamefont {Katzgraber}\ \emph {et~al.}(2006)\citenamefont
  {Katzgraber}, \citenamefont {Trebst}, \citenamefont {Huse},\ and\
  \citenamefont {Troyer}}]{katzgraber06jsm}%
  \BibitemOpen
  \bibfield  {author} {\bibinfo {author} {\bibfnamefont {H.~G.}\ \bibnamefont
  {Katzgraber}}, \bibinfo {author} {\bibfnamefont {S.}~\bibnamefont {Trebst}},
  \bibinfo {author} {\bibfnamefont {D.~A.}\ \bibnamefont {Huse}}, \ and\
  \bibinfo {author} {\bibfnamefont {M.}~\bibnamefont {Troyer}},\ }\href
  {\doibase 10.1088/1742-5468/2006/03/P03018} {\bibfield  {journal} {\bibinfo
  {journal} {J. Stat. Mech.: Theory Exp.}\ }\textbf {\bibinfo {volume}
  {2006}},\ \bibinfo {pages} {P03018} (\bibinfo {year} {2006})}\BibitemShut
  {NoStop}%
\bibitem [{\citenamefont {Trebst}\ \emph {et~al.}(2006)\citenamefont {Trebst},
  \citenamefont {Troyer},\ and\ \citenamefont {Hansmann}}]{trebst06jcp}%
  \BibitemOpen
  \bibfield  {author} {\bibinfo {author} {\bibfnamefont {S.}~\bibnamefont
  {Trebst}}, \bibinfo {author} {\bibfnamefont {M.}~\bibnamefont {Troyer}}, \
  and\ \bibinfo {author} {\bibfnamefont {U.~H.~E.}\ \bibnamefont {Hansmann}},\
  }\href {\doibase http://dx.doi.org/10.1063/1.2186639} {\bibfield  {journal}
  {\bibinfo  {journal} {J. Chem. Phys.}\ }\textbf {\bibinfo {volume} {124}},\
  \bibinfo {pages} {174903} (\bibinfo {year} {2006})}\BibitemShut {NoStop}%
\bibitem [{\citenamefont {Neuhaus}\ \emph {et~al.}(2007)\citenamefont
  {Neuhaus}, \citenamefont {Magiera},\ and\ \citenamefont
  {Hansmann}}]{neuhaus07pre}%
  \BibitemOpen
  \bibfield  {author} {\bibinfo {author} {\bibfnamefont {T.}~\bibnamefont
  {Neuhaus}}, \bibinfo {author} {\bibfnamefont {M.~P.}\ \bibnamefont
  {Magiera}}, \ and\ \bibinfo {author} {\bibfnamefont {U.~H.~E.}\ \bibnamefont
  {Hansmann}},\ }\href {\doibase 10.1103/PhysRevE.76.045701} {\bibfield
  {journal} {\bibinfo  {journal} {Phys. Rev. E}\ }\textbf {\bibinfo {volume}
  {76}},\ \bibinfo {pages} {045701} (\bibinfo {year} {2007})}\BibitemShut
  {NoStop}%
\bibitem [{\citenamefont {Patriksson}\ and\ \citenamefont {van~der
  Spoel}(2008)}]{patriksson08pccp}%
  \BibitemOpen
  \bibfield  {author} {\bibinfo {author} {\bibfnamefont {A.}~\bibnamefont
  {Patriksson}}\ and\ \bibinfo {author} {\bibfnamefont {D.}~\bibnamefont
  {van~der Spoel}},\ }\href {\doibase 10.1039/B716554D} {\bibfield  {journal}
  {\bibinfo  {journal} {Phys. Chem. Chem. Phys.}\ }\textbf {\bibinfo {volume}
  {10}},\ \bibinfo {pages} {2073} (\bibinfo {year} {2008})}\BibitemShut
  {NoStop}%
\bibitem [{\citenamefont {Bittner}\ \emph {et~al.}(2008)\citenamefont
  {Bittner}, \citenamefont {Nu\ss{}baumer},\ and\ \citenamefont
  {Janke}}]{bittner08prl}%
  \BibitemOpen
  \bibfield  {author} {\bibinfo {author} {\bibfnamefont {E.}~\bibnamefont
  {Bittner}}, \bibinfo {author} {\bibfnamefont {A.}~\bibnamefont
  {Nu\ss{}baumer}}, \ and\ \bibinfo {author} {\bibfnamefont {W.}~\bibnamefont
  {Janke}},\ }\href {\doibase 10.1103/PhysRevLett.101.130603} {\bibfield
  {journal} {\bibinfo  {journal} {Phys. Rev. Lett.}\ }\textbf {\bibinfo
  {volume} {101}},\ \bibinfo {pages} {130603} (\bibinfo {year}
  {2008})}\BibitemShut {NoStop}%
\bibitem [{\citenamefont {Ballard}\ and\ \citenamefont
  {Wales}(2014)}]{ballard14jctc}%
  \BibitemOpen
  \bibfield  {author} {\bibinfo {author} {\bibfnamefont {A.~J.}\ \bibnamefont
  {Ballard}}\ and\ \bibinfo {author} {\bibfnamefont {D.~J.}\ \bibnamefont
  {Wales}},\ }\href {\doibase 10.1021/ct500797a} {\bibfield  {journal}
  {\bibinfo  {journal} {J. Chem. Theory Comput. (JCTC)}\ }\textbf {\bibinfo
  {volume} {10}},\ \bibinfo {pages} {5599} (\bibinfo {year}
  {2014})}\BibitemShut {NoStop}%
\bibitem [{\citenamefont {Williams}\ \emph {et~al.}(2006)\citenamefont
  {Williams}, \citenamefont {Mishin},\ and\ \citenamefont
  {Hamilton}}]{williams06msmse}%
  \BibitemOpen
  \bibfield  {author} {\bibinfo {author} {\bibfnamefont {P.~L.}\ \bibnamefont
  {Williams}}, \bibinfo {author} {\bibfnamefont {Y.}~\bibnamefont {Mishin}}, \
  and\ \bibinfo {author} {\bibfnamefont {J.~C.}\ \bibnamefont {Hamilton}},\
  }\href {\doibase 10.1088/0965-0393/14/5/002} {\bibfield  {journal} {\bibinfo
  {journal} {Model. Simul. Mater. Sci. Eng.}\ }\textbf {\bibinfo {volume}
  {14}},\ \bibinfo {pages} {817} (\bibinfo {year} {2006})}\BibitemShut
  {NoStop}%
\bibitem [{\citenamefont {Xiao}\ and\ \citenamefont {Hu}(2006)}]{xiao06jcp}%
  \BibitemOpen
  \bibfield  {author} {\bibinfo {author} {\bibfnamefont {S.}~\bibnamefont
  {Xiao}}\ and\ \bibinfo {author} {\bibfnamefont {W.}~\bibnamefont {Hu}},\
  }\href {\doibase 10.1063/1.2209227} {\bibfield  {journal} {\bibinfo
  {journal} {J. Chem. Phys.}\ }\textbf {\bibinfo {volume} {125}},\ \bibinfo
  {eid} {014503} (\bibinfo {year} {2006})}\BibitemShut {NoStop}%
\bibitem [{\citenamefont {Tian}\ \emph {et~al.}(2008)\citenamefont {Tian},
  \citenamefont {Liu}, \citenamefont {Liu}, \citenamefont {Zheng},
  \citenamefont {Hou},\ and\ \citenamefont {Peng}}]{Tian08ncs}%
  \BibitemOpen
  \bibfield  {author} {\bibinfo {author} {\bibfnamefont {Z.-A.}\ \bibnamefont
  {Tian}}, \bibinfo {author} {\bibfnamefont {R.-S.}\ \bibnamefont {Liu}},
  \bibinfo {author} {\bibfnamefont {H.-R.}\ \bibnamefont {Liu}}, \bibinfo
  {author} {\bibfnamefont {C.-X.}\ \bibnamefont {Zheng}}, \bibinfo {author}
  {\bibfnamefont {Z.-Y.}\ \bibnamefont {Hou}}, \ and\ \bibinfo {author}
  {\bibfnamefont {P.}~\bibnamefont {Peng}},\ }\href {\doibase
  10.1016/j.jnoncrysol.2008.04.006} {\bibfield  {journal} {\bibinfo  {journal}
  {J. Non-Cryst. Solids}\ }\textbf {\bibinfo {volume} {354}},\ \bibinfo {pages}
  {3705} (\bibinfo {year} {2008})}\BibitemShut {NoStop}%
\bibitem [{\citenamefont {Jian}\ \emph {et~al.}(2010)\citenamefont {Jian},
  \citenamefont {Chen}, \citenamefont {Chang}, \citenamefont {Zeng},
  \citenamefont {He},\ and\ \citenamefont {Jie}}]{jian10scts}%
  \BibitemOpen
  \bibfield  {author} {\bibinfo {author} {\bibfnamefont {Z.}~\bibnamefont
  {Jian}}, \bibinfo {author} {\bibfnamefont {J.}~\bibnamefont {Chen}}, \bibinfo
  {author} {\bibfnamefont {F.}~\bibnamefont {Chang}}, \bibinfo {author}
  {\bibfnamefont {Z.}~\bibnamefont {Zeng}}, \bibinfo {author} {\bibfnamefont
  {T.}~\bibnamefont {He}}, \ and\ \bibinfo {author} {\bibfnamefont
  {W.}~\bibnamefont {Jie}},\ }\href {\doibase 10.1007/s11431-010-4171-5}
  {\bibfield  {journal} {\bibinfo  {journal} {Sci. China Technol. Sci.}\
  }\textbf {\bibinfo {volume} {53}},\ \bibinfo {pages} {3203} (\bibinfo {year}
  {2010})}\BibitemShut {NoStop}%
\bibitem [{\citenamefont {Liu}\ \emph {et~al.}(2001)\citenamefont {Liu},
  \citenamefont {Xia}, \citenamefont {Zhu},\ and\ \citenamefont
  {Sun}}]{liu01jcp}%
  \BibitemOpen
  \bibfield  {author} {\bibinfo {author} {\bibfnamefont {C.~S.}\ \bibnamefont
  {Liu}}, \bibinfo {author} {\bibfnamefont {J.}~\bibnamefont {Xia}}, \bibinfo
  {author} {\bibfnamefont {Z.~G.}\ \bibnamefont {Zhu}}, \ and\ \bibinfo
  {author} {\bibfnamefont {D.~Y.}\ \bibnamefont {Sun}},\ }\href {\doibase
  10.1063/1.1362292} {\bibfield  {journal} {\bibinfo  {journal} {J. Chem.
  Phys.}\ }\textbf {\bibinfo {volume} {114}},\ \bibinfo {pages} {7506}
  (\bibinfo {year} {2001})}\BibitemShut {NoStop}%
\bibitem [{\citenamefont {Li}\ \emph {et~al.}(2011)\citenamefont {Li},
  \citenamefont {Liu}, \citenamefont {Hou}, \citenamefont {Chen},\ and\
  \citenamefont {Chen}}]{li2011int}%
  \BibitemOpen
  \bibfield  {author} {\bibinfo {author} {\bibfnamefont {F.}~\bibnamefont
  {Li}}, \bibinfo {author} {\bibfnamefont {X.}~\bibnamefont {Liu}}, \bibinfo
  {author} {\bibfnamefont {H.}~\bibnamefont {Hou}}, \bibinfo {author}
  {\bibfnamefont {G.}~\bibnamefont {Chen}}, \ and\ \bibinfo {author}
  {\bibfnamefont {G.}~\bibnamefont {Chen}},\ }\href {\doibase
  10.1016/j.intermet.2010.12.016} {\bibfield  {journal} {\bibinfo  {journal}
  {Intermetallics}\ }\textbf {\bibinfo {volume} {19}},\ \bibinfo {pages} {630}
  (\bibinfo {year} {2011})}\BibitemShut {NoStop}%
\bibitem [{\citenamefont {Kim}\ \emph {et~al.}(2010)\citenamefont {Kim},
  \citenamefont {Keyes},\ and\ \citenamefont {Straub}}]{kim10jcp}%
  \BibitemOpen
  \bibfield  {author} {\bibinfo {author} {\bibfnamefont {J.}~\bibnamefont
  {Kim}}, \bibinfo {author} {\bibfnamefont {T.}~\bibnamefont {Keyes}}, \ and\
  \bibinfo {author} {\bibfnamefont {J.~E.}\ \bibnamefont {Straub}},\ }\href
  {\doibase 10.1063/1.3432176} {\bibfield  {journal} {\bibinfo  {journal} {J.
  Chem. Phys.}\ }\textbf {\bibinfo {volume} {132}},\ \bibinfo {pages} {224107}
  (\bibinfo {year} {2010})}\BibitemShut {NoStop}%
\bibitem [{\citenamefont {Sugita}\ and\ \citenamefont
  {Okamoto}(2000)}]{sugita00cpl}%
  \BibitemOpen
  \bibfield  {author} {\bibinfo {author} {\bibfnamefont {Y.}~\bibnamefont
  {Sugita}}\ and\ \bibinfo {author} {\bibfnamefont {Y.}~\bibnamefont
  {Okamoto}},\ }\href {\doibase 10.1016/S0009-2614(00)00999-4} {\bibfield
  {journal} {\bibinfo  {journal} {Chem. Phys. Lett.}\ }\textbf {\bibinfo
  {volume} {329}},\ \bibinfo {pages} {261} (\bibinfo {year}
  {2000})}\BibitemShut {NoStop}%
\bibitem [{\citenamefont {Kim}\ \emph {et~al.}(2012)\citenamefont {Kim},
  \citenamefont {Straub},\ and\ \citenamefont {Keyes}}]{kim12jpcb}%
  \BibitemOpen
  \bibfield  {author} {\bibinfo {author} {\bibfnamefont {J.}~\bibnamefont
  {Kim}}, \bibinfo {author} {\bibfnamefont {J.~E.}\ \bibnamefont {Straub}}, \
  and\ \bibinfo {author} {\bibfnamefont {T.}~\bibnamefont {Keyes}},\ }\href
  {\doibase 10.1021/jp300366j} {\bibfield  {journal} {\bibinfo  {journal} {J.
  Phys. Chem. B}\ }\textbf {\bibinfo {volume} {116}},\ \bibinfo {pages} {8646}
  (\bibinfo {year} {2012})}\BibitemShut {NoStop}%
\bibitem [{\citenamefont {Vogel}\ \emph {et~al.}(2013)\citenamefont {Vogel},
  \citenamefont {Li}, \citenamefont {W\"ust},\ and\ \citenamefont
  {Landau}}]{vogel13prl}%
  \BibitemOpen
  \bibfield  {author} {\bibinfo {author} {\bibfnamefont {T.}~\bibnamefont
  {Vogel}}, \bibinfo {author} {\bibfnamefont {Y.~W.}\ \bibnamefont {Li}},
  \bibinfo {author} {\bibfnamefont {T.}~\bibnamefont {W\"ust}}, \ and\ \bibinfo
  {author} {\bibfnamefont {D.~P.}\ \bibnamefont {Landau}},\ }\href {\doibase
  10.1103/PhysRevLett.110.210603} {\bibfield  {journal} {\bibinfo  {journal}
  {Phys. Rev. Lett.}\ }\textbf {\bibinfo {volume} {110}},\ \bibinfo {pages}
  {210603} (\bibinfo {year} {2013})}\BibitemShut {NoStop}%
\bibitem [{\citenamefont {Vogel}\ \emph {et~al.}(2014)\citenamefont {Vogel},
  \citenamefont {Li}, \citenamefont {W\"ust},\ and\ \citenamefont
  {Landau}}]{vogel14pre}%
  \BibitemOpen
  \bibfield  {author} {\bibinfo {author} {\bibfnamefont {T.}~\bibnamefont
  {Vogel}}, \bibinfo {author} {\bibfnamefont {Y.~W.}\ \bibnamefont {Li}},
  \bibinfo {author} {\bibfnamefont {T.}~\bibnamefont {W\"ust}}, \ and\ \bibinfo
  {author} {\bibfnamefont {D.~P.}\ \bibnamefont {Landau}},\ }\href {\doibase
  10.1103/PhysRevE.90.023302} {\bibfield  {journal} {\bibinfo  {journal} {Phys.
  Rev. E}\ }\textbf {\bibinfo {volume} {90}},\ \bibinfo {pages} {023302}
  (\bibinfo {year} {2014})}\BibitemShut {NoStop}%
\bibitem [{\citenamefont {Allen}\ and\ \citenamefont
  {Quigley}(2013)}]{allen13molphys}%
  \BibitemOpen
  \bibfield  {author} {\bibinfo {author} {\bibfnamefont {M.~P.}\ \bibnamefont
  {Allen}}\ and\ \bibinfo {author} {\bibfnamefont {D.}~\bibnamefont
  {Quigley}},\ }\href {\doibase 10.1080/00268976.2013.817623} {\bibfield
  {journal} {\bibinfo  {journal} {Mol. Phys.}\ }\textbf {\bibinfo {volume}
  {111}},\ \bibinfo {pages} {3442} (\bibinfo {year} {2013})}\BibitemShut
  {NoStop}%
\bibitem [{\citenamefont {Junghans}\ \emph {et~al.}(2006)\citenamefont
  {Junghans}, \citenamefont {Bachmann},\ and\ \citenamefont
  {Janke}}]{junghans06prl}%
  \BibitemOpen
  \bibfield  {author} {\bibinfo {author} {\bibfnamefont {C.}~\bibnamefont
  {Junghans}}, \bibinfo {author} {\bibfnamefont {M.}~\bibnamefont {Bachmann}},
  \ and\ \bibinfo {author} {\bibfnamefont {W.}~\bibnamefont {Janke}},\ }\href
  {\doibase 10.1103/PhysRevLett.97.218103} {\bibfield  {journal} {\bibinfo
  {journal} {Phys. Rev. Lett.}\ }\textbf {\bibinfo {volume} {97}},\ \bibinfo
  {pages} {218103} (\bibinfo {year} {2006})}\BibitemShut {NoStop}%
\bibitem [{\citenamefont {Schnabel}\ \emph {et~al.}(2011)\citenamefont
  {Schnabel}, \citenamefont {Seaton}, \citenamefont {Landau},\ and\
  \citenamefont {Bachmann}}]{schnabel11pre}%
  \BibitemOpen
  \bibfield  {author} {\bibinfo {author} {\bibfnamefont {S.}~\bibnamefont
  {Schnabel}}, \bibinfo {author} {\bibfnamefont {D.~T.}\ \bibnamefont
  {Seaton}}, \bibinfo {author} {\bibfnamefont {D.~P.}\ \bibnamefont {Landau}},
  \ and\ \bibinfo {author} {\bibfnamefont {M.}~\bibnamefont {Bachmann}},\
  }\href {\doibase 10.1103/PhysRevE.84.011127} {\bibfield  {journal} {\bibinfo
  {journal} {Phys. Rev. E}\ }\textbf {\bibinfo {volume} {84}},\ \bibinfo
  {pages} {011127} (\bibinfo {year} {2011})}\BibitemShut {NoStop}%
\bibitem [{\citenamefont {Burden}\ and\ \citenamefont
  {Faires}(1985)}]{burden85book}%
  \BibitemOpen
  \bibfield  {author} {\bibinfo {author} {\bibfnamefont {R.~L.}\ \bibnamefont
  {Burden}}\ and\ \bibinfo {author} {\bibfnamefont {J.~D.}\ \bibnamefont
  {Faires}},\ }\href@noop {} {\emph {\bibinfo {title} {Numerical Analysis}}},\
  \bibinfo {edition} {3rd}\ ed.\ (\bibinfo  {publisher} {PWS Publishers},\
  \bibinfo {year} {1985})\BibitemShut {NoStop}%
\end{thebibliography}%

\section*{Supplemental Material}

\subsection*{Microcanonical temperature measurements}

In this work we rely on methods that use measurements of
temperature-like observables in order to estimate the derivative of
the entropy (or, equivalently, of the density of states) with respect
to energy.  For example, the microcanonical temperature can be
written in terms of configurational properties as
\begin{equation}
 \frac{1}{k_B T_\mathrm{conf}(U)}=
   \left\langle \nabla \cdot \left( \frac{\nabla U}{\nabla
         U\cdot\nabla U} \right) \right\rangle_U\,
\label{eq:SM_1}
\end{equation}
where $\langle\ldots\rangle_U$ is the average over states with a given
potential energy $U$~\cite{rugh97prl,butler98jcp}. For canonical
averages or for very large system sizes, $T_\mathrm{conf}(U)$,
measured from configurations in a heat-bath at constant temperature,
is equal to just that canonical
temperature~\cite{butler98jcp,allen13molphys}.  Another way of
estimating $\partial S/\partial U$ is based on thermodynamic
integration using time derivatives, as derived in the context of the
adaptive biasing force (ABF) method~\cite{darve08jcp}.  By using the
potential energy as the 'order parameter', $T_\mathrm{m}(U)=(\partial
S(U)/\partial U)^{-1}$ can be estimated via
\begin{equation}
  T_\mathrm{ABF}(U)=T_0/(1-s(U))\,,
\label{eq:SM_2}
\end{equation}
where $T_0$ is the heat-bath temperature of the particular
canonical-ensemble MD walker and with
\begin{equation}
  s(U)=-\left\langle \frac{\mathrm{d}}{\mathrm{d}t}
    \left(\frac{\nabla U}{\nabla U\cdot \nabla U}\right)\cdot p\right\rangle_U\,, 
\end{equation}
where $p$ are the particle momenta. In contrast to
Eq.~(\ref{eq:SM_1}), we do not need to calculate second derivatives of
the potential energy, but only a time-derivative obtained in
finite-difference form using a constant-energy MD timestep. This can
be much less costly, especially for a complex many-body potentials,
such as the one used in our work. However, if $p$ are not available,
in a MC simulation, for example, one can instead use
$T_\mathrm{conf}(U)$. In Fig.~\ref{fig_SM1} we show raw measurements
$T^\prime$ that contribute to the averages in Eqs.~(\ref{eq:SM_1})
and~(\ref{eq:SM_2}).  Every data point corresponds to measurements of
$T^\prime_\mathrm{ABF}$ and $T^\prime_\mathrm{conf}$ for one
particular microstate sampled during the run. An extremely good
agreement can be seen for both observables, not only with respect to
each other but also with respect to the canonical reference
temperatures indicated by the dotted lines.  After carrying out the
averages, we do not see any difference in the results of our
simulations when using either of the observables and make our choice
solely based on the computational performance.

As a valuable side-product, note that since $T_\mathrm{m}(U)$ is
\emph{measured}, the results are readily available for a
microcanonical analysis of finite-size phase
transitions~\cite{junghans06prl,schnabel11pre}.

\begin{figure}
\includegraphics[width=\columnwidth]{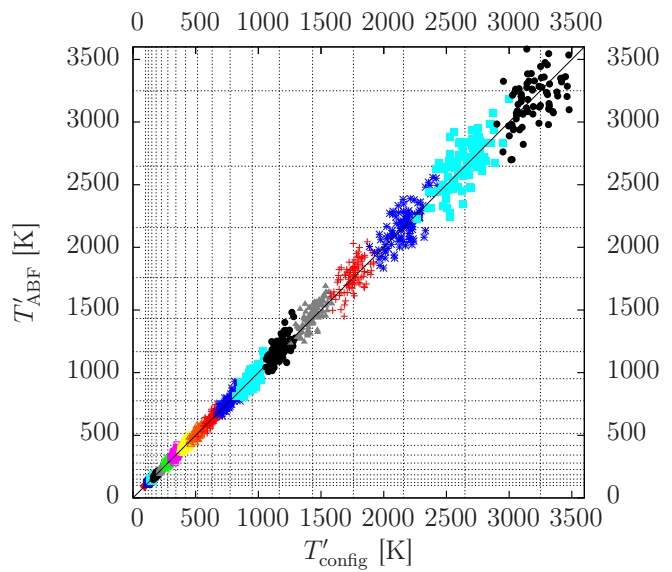}
\caption{$T_\mathrm{config}$ vs. $T_\mathrm{ABF}$ for one example g-RE run
   with a fixed temperature set for the
  $N=500$ Ag system. Black, diagonal line: $T_\mathrm{config}=T_\mathrm{ABF}$ to guide the
  eye; Dotted lines: individual reference temperatures $T_i$. See text for details.
  \label{fig_SM1}}
\end{figure}

\subsection*{Determination of temperature sets}

The transition (exchange) probability between two energy distributions
$P_{i+1}$ and $P_i$ corresponding to different temperatures $\beta_i$
and $\beta_{i+1}$ takes the form
\begin{align}
W^\mathrm{tot}_\mathrm{acc}(P_{i+1}, P_i)=\!\!\iint_{-\infty}^\infty\!\!
P_{i}(U) W_\mathrm{acc}(U,U^\prime)
P_{{i+1}}(U^\prime) \, \mathrm{d}U \mathrm{d}U^\prime.
\label{eq:SM_3}
\end{align}
$W_\mathrm{acc}(U,U^\prime)$ is the usual acceptance probability for a
replica exchange in canonical parallel tempering simulations:
$W_\mathrm{acc}(U,U^\prime)=\min[1,\mathrm{e}^{\Delta\beta \Delta
  U]}$, with $\Delta\beta$ being the difference in inverse temperature
and $\Delta U$ the energy difference of the corresponding replicas.
The integral in Eq.~(\ref{eq:SM_3}) is evaluated numerically as a sum
over histogram bins. Recalling that $g(U)$ is used to calculate the energy
distributions $P_i(U)=g(U)\exp[-\beta_iU]$, it is clear that the
integrand takes non-negligible values only in a limited energy range,
and hence that the bounds of the integrals can be restricted in practice.  For
example, the estimator of $g(U)$ is zero (by construction) at energies
lower than the lowest energy observed by any replica, hence so is
$P_i(U)=g(U)\exp[-\beta_iU]$. Similarly, at high energies, the
Boltzmann factor leads to a rapid decay of $P_i(U)$, so that the
integral can be truncated when $P_i(U)$ becomes negligible. 

To find the optimal
temperature set, we first (arbitrarily) select a target exchange
probability $c_t$ and set $\beta_1=1/k_B T_\mathrm{min}$. 
We then
successively apply a binary search (also known as
interval halving method or bisection method; a standard method in
numerical analysis~\cite{burden85book}) to neighboring pairs of
walkers to find $\beta_{i+1}$ such that
$W^\mathrm{tot}_\mathrm{acc}(P_{i+1},P_i)=c_t$. In practice, we first
bracket the interval of $\beta_{i+1}$ that contains the solution to
$W^\mathrm{tot}_\mathrm{acc}(P_{i+1},P_i)=c_t$,
and then apply a conventional binary search to identify the precise value.
Given that the number $N$ of replicas
is constant, a given value of $c_t$ implies a given value of $T_N$.
To ensure that we always cover the same temperature range, we then
apply an additional bisection scheme where we determine $c_t$ such
that $T_N$ falls within a small range around the targeted maximal
temperature $T_\mathrm{max}$. That is, whenever $T_N$ turns out to be
too small we decrease $c_t$ (and vice versa), in ever smaller
amounts~\cite{burden85book} and repeat the procedure described above.
Of course, if the converged $c_t$ is too small, one should increase
the number of replicas $N$.  Note that Eq.~(2) (in the main article)
only needs to be satisfied to some moderate precision to reap most of
the performance benefits. Therefore, the tolerance in the bisection
scheme, i.e., the range around the target value which is acceptable,
can be adjusted. In all our simulations we observed that the
determination of the optimal temperature set required only a
negligible amount of computing time compared to the cost of actual MD,
even though one evaluates a double-integral within a double-bisection
scheme.


To emphasize the robustness of our scheme, we choose a
far-from-optimal initial set $\{T_i\}$ for the run shown in Fig.~1 in
the main article. About half of the walkers start at $T_\mathrm{min}$,
the other half at $T_\mathrm{max}$; where there is, for all practical
purposes, no overlap at all between the corresponding canonical
distributions. Furthermore, recall that all walkers start with the
same initial configuration.  Still, the temperature set spreads out
rapidly (see Fig.~\ref{fig_SM2}, in analogy to corresponding figures
in Refs.~\cite{trebst06jcp,katzgraber06jsm}) so that the walkers cover
the whole range of temperatures and energies after a few adaptation
steps. This illustrates that the method does not depend on the
availability of any prior information on the system -- in particular,
no assumption or pre-estimation of the temperature set is necessary
 -- and that it is very robust against even extreme choices of
initial conditions. 
\begin{figure}
\includegraphics[width=\columnwidth]{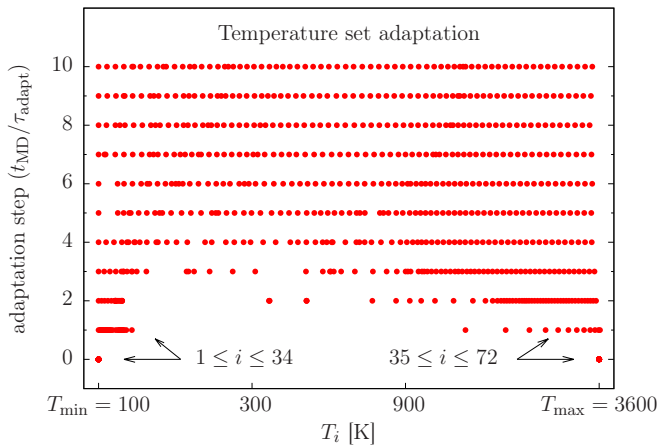}
\caption{Convergence of the temperature set $\{T_i\}$ after an initialization where
  every walker starts either at the lower or the upper end of the total temperature range. The plot  
  shows the evolution of $\{T_i\}$ during the first ten
  adaptation steps; after very few steps, the simulation optimally covers the whole 
  temperature range. Data measured during the same simulation run
  shown in Fig.~1 in the main article. 
  \label{fig_SM2}}
\end{figure}

\subsection*{Convergence to the canonical distribution}
Despite of the fact that the g-RE procedure is history-dependent,
i.e., that the result of the resampling depends on the previous
history of the simulation through the content of the database, the method nonetheless converges to the
canonical distribution. This can be seen by first considering a
simpler situation: a single walker evolving through a canonical
sampler and a database resampler. In this case, our database resampling procedure becomes:

\begin{itemize}
\item
Sample a new configuration from the database
\item
Propagate for some time $\Delta t$ with a canonical sampler
\item
Insert the current configuration back into the database
\item
If the number of configuration in the database exceeds $M$, remove an
element at random. Equivalently, old configurations could be purged
from the database after a set time (as done in the manuscript).
\item
Repeat
\end{itemize}

Following this prescription, after $N$ cycles of the algorithm, every
member of the database is the product of a proper time evolution.
Indeed, from the perspective of individual configurations, they are
randomly sampled, propagated for some time t with a canonical sampler,
and then replaced in the database, i.e., {\em they are the product of
  an unbiased sequence of propagation with the correct dynamics}.
However, because the configurations in the database are correlated,
resampling from the {\em same} database does not produce independent
canonical samples.  For example, consider the case $M=2$: both
configurations are now likely to be extremely similar if the
propagation time $\Delta t$ is short. As we do not require
independence, this is not an issue for g-RE in practice.

The second crucial point is that the ``age'' of the configurations in
the database, i.e., the total (aggregate) amount of time they were
propagated for, scales with the total simulation time. Therefore, in
spite of correlations between individual entries, the sampling from
the database over long times will be correct. Introducing a memory
time after which configurations are dropped, as we did in the
manuscript, directly imposes such a condition: ``young''
configurations cannot persist in the database by construction.
Numerical simulations of the random deletion procedure demonstrate
that this is also true there, to a remarkable extend. In fact, the
mean age of a configuration is $\sim N \Delta t/M$, as can be
intuitively expected. What is perhaps surprising is that the standard
deviation of the age is a time ($N$)-independent constant that scales as
$M^{1/2}$, i.e., the relative fluctuation (standard deviation of the
age over the average age) {\em vanishes} at large $N$. Results for a
memory size of 100 are shown in Fig.~\ref{fig_SM3}.
\begin{figure}
\includegraphics[width=\columnwidth]{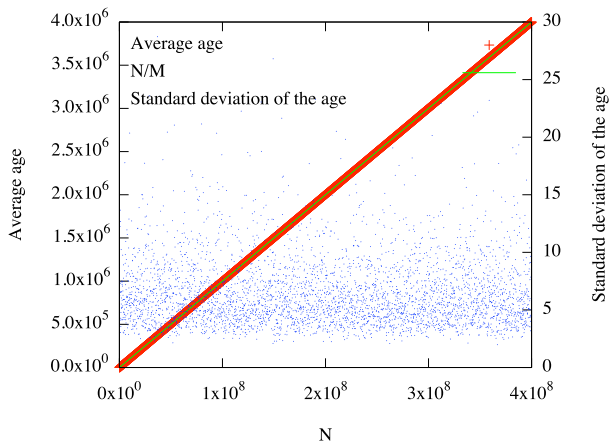}
\caption{Statistics of the age of configurations in the database vs. $N$ for $M=100$.
\label{fig_SM3}
}
\end{figure}
\begin{figure}
\includegraphics[width=\columnwidth]{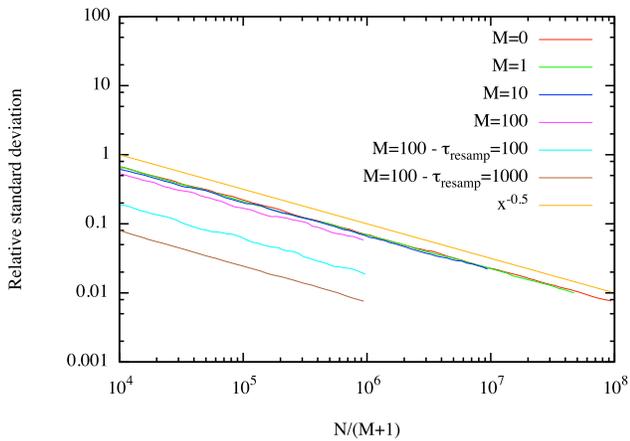}
\caption{Relative standard deviation (w.r.t. the average) of the position $x$ of a walker for various memory size and resampling times. Unless otherwise specified, the resampling time is 1. 
\label{fig_SM4}
}
\end{figure}

This shows that, while the average age of each configuration in the
database is smaller than $N$, it scales with simulation time, and it is
exceedingly unlikely that a very ``young'' configuration will be
sampled. Taken together with the fact that each one of them is the
product of a proper time evolution, this shows the above procedure
yield a canonical sampler over long enough times. In the worst case
(when $\Delta t$ is small) the configurations in the database will
form a tight bundle in phase space. However, the bundle itself will
evolve according to proper dynamics. Therefore, sampling from the
database will yield unbiased canonical results over long times, so
that we have an equal probability of sampling from the set of states
with the same potential energy. As discussed in the manuscript, this
is the only requirement for g-RE to be a valid sampler. However one
should not expect multiple samples obtained from the database {\em at
  the same time } to be independent.  Note that the same argument
translates directly to the RE case, i.e., each configuration in the
database taken alone, is a the product of proper dynamics whose age
scales with the simulation time.
We note that the situation appears problematic in the infinite $M$ limit
or in the infinite memory time limit. As this is not a situation we
have to confront in practice (because of finite computer memory), we
do not explore that limit further.

The above discussion has been demonstrated using direct numerical
simulations. We now consider a two-replica case. For simplicity, the
system is a 1D square well in $0<x<1$. Given the flat energy
landscape, the temperature is irrelevant, so doing RE would yield
acceptance with unit probability and thus would yield identical
results. The dynamics are carried out with Metropolis MC with a
Gaussian proposal distribution of standard deviation 0.01. Unless
otherwise noted, we carry out only one MC step between database
resampling ($\tau_\mathrm{resamp}=1$). In Fig.~\ref{fig_SM4}, we
report the relative standard deviation (the standard deviation over
the average) of an histogram of the visited positions of one of the
walkers as a function of $N/(M+1)$ for $M=0$ (standard MC), $1$, $10,$
and $100$. Note that each walker populates its own database but can
only sample from the database of the other walker (which is equivalent
to the setup used in the manuscript).  Since the canonical
distribution is uniform, convergence is indicated by a vanishing
relative deviation at long times. As shown in the figure, all cases
converge at roughly the same rate once the average age in the database
is accounted for.  Note that, because of decreased correlations due to
the larger database, the $M=100$ case is converging even faster than
expected from simple scaling arguments. Indeed, running the same case
but resampling only every 100 or 1000 MC steps, thereby further
limiting the correlation between entries in the database, yields an
even faster convergence.  Eventually, in the limit where the
resampling time exceeds the ergodic relaxation time of the dynamics,
the standard MC results (vs. $N$, not $N/M$ as plotted) are recovered,
which is the expected behavior.

\end{document}